# Silicon detectors: damage, modelling and expected long-time behaviour in physics experiments at ultra high energy

Ionel Lazanu,[a] Sorina Lazanu[b]

[a]University of Bucharest, Faculty of Physics, POBox MG-11, Bucharest-Magurele, Romania

[b]National Institute for Materials Physics, POBox MG-7, Bucharest-Magurele, Romania

**Abstract**

In this contribution, the structural modifications of the material and the degradation of devices is modelled and compared with experimental data for more resistivities, temperatures, crystal orientations and oxygen concentrations, considering the existence of the new primary fourfold coordinated defect, besides the vacancy and the interstitial. Some estimations of the behaviour of detectors in concrete environments at the next generations of high energy physics experiments as LHC, SLHC, VLHC, or ULHC are done.

Bulk displacement damage in silicon produces effects at the device level and limits their long time utilisation in detector systems. So, the modelling of the damage produced by the radiation fields in silicon and in silicon detectors, and the prediction of the time behaviour of detectors in different hostile environments, represents a useful tool which will permit taking the best decisions for obtaining radiation harder materials.

In the present contribution, starting from the interaction of the incoming particle with the lattice [1], with the formation of primary defects, where the existence and characteristics of new $Si_{FFCD}$ defect is considered [2, 3], the kinetics of defects is treated in the frame of a model based on the theory of diffusion limited reactions, developed previously by the authors [4]. The effects on device parameters are investigated.

The time dependence of the degradation of different silicon materials which differ by resistivity, crystal orientation, oxygen concentration and temperature of irradiation and operation after proton irradiation is obtained and presented in Figures 1 and 2. The predictions of the model are compared with experimental available data [5, 6], regarding the modifications produced: increase of leakage current and change of effective carrier concentration.

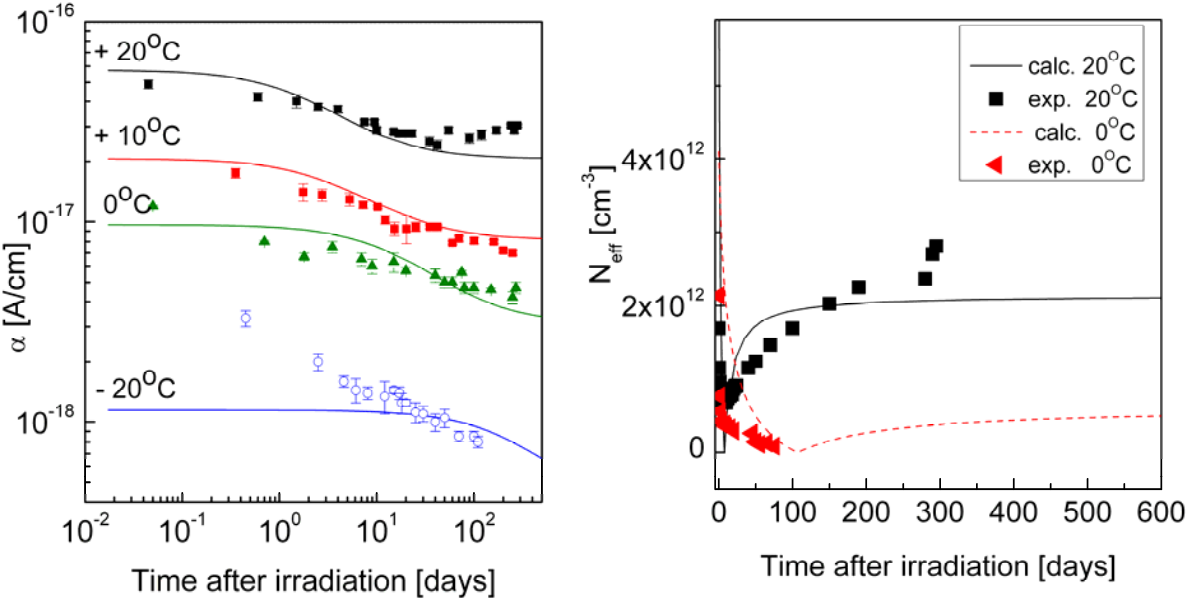

Fig. 1.
Time dependence of the α constant of the leakage current (a) and of $N_{eff}$ at $0\,^{\circ}C$ and $20\,^{\circ}C$ (b).

The anisotropy of the effective threshold energy in silicon at irradiation after different crystal directions is clearly observed in FZ materials and is ambiguous in DOFZ and this behaviour is tentatively attributed to the deformation of lattice due to the increase of oxygen concentration. The theoretical predictions of threshold energies in silicon are controversial – see the review in Ref. [7], in conditions of lack of systematic experimental studies.

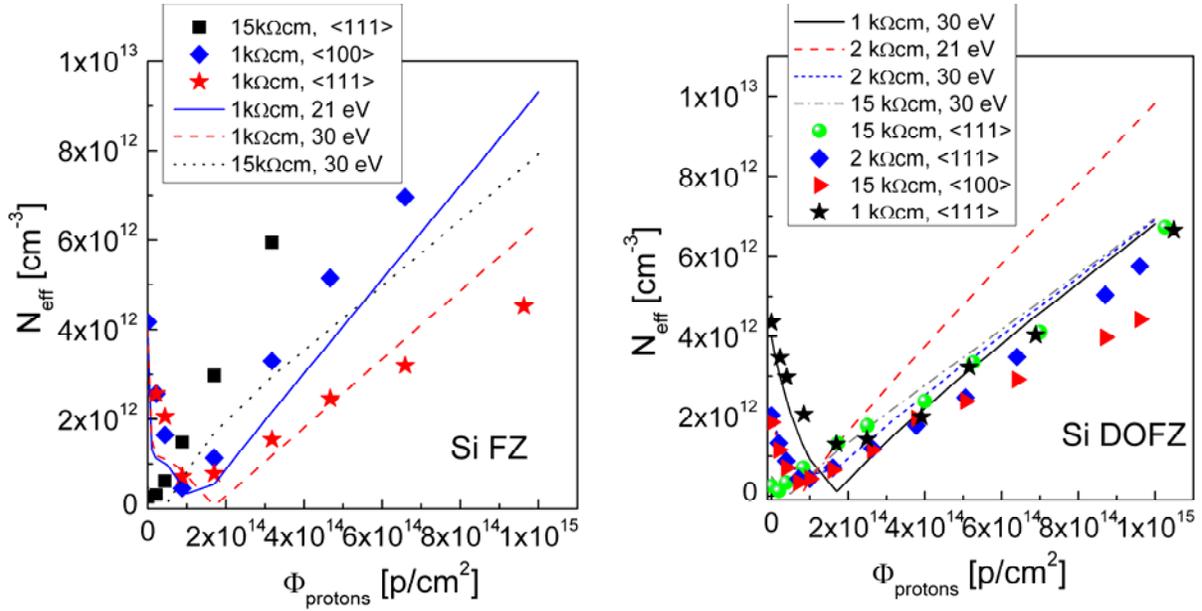

Fig. 2.
Fluence dependence of $N_{eff}$ for detectors manufactured from different materials.

The model reproduces well or reasonably well the time behaviour of detectors with different resistivities and oxygen contents between 20 $^{0}$C and -20 $^{0}$C after proton irradiation at fluences up to $10^{15}$ part/cm$^2$.

The long time behaviour of silicon detectors in some ultra high energy radiation environments is predicted. Scenarios for the next SLHC, VLHC, ULHC colliders are considered in agreement with the hypothesis mentioned in Ref. [8].

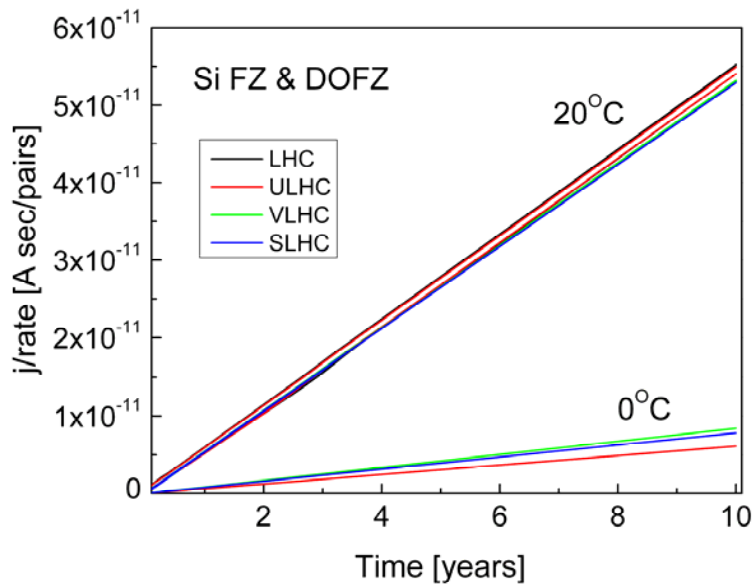

Fig. 3.
Time dependence of the volume density of the leakage current normalised to the generation rate, in different environments, at 0 $^0$C and 20 $^0$C.

An interesting model prediction is that the leakage current scales with the rate of generation of primary defects and is roughly independent on particle spectra and material technology (FZ or DOFZ), and the rate of increase of the current is about 5x10$^{-12}$ Asec/pair/year at 20 $^0$C and 8x10$^{-13}$ Asec/pair/year at 0 $^0$C – see Figure 3.

In Figure 4, the time variation of $N_{eff}$ at 0 $^0$C suggests that DOFZ silicon is radiation harder in respect to FZ.

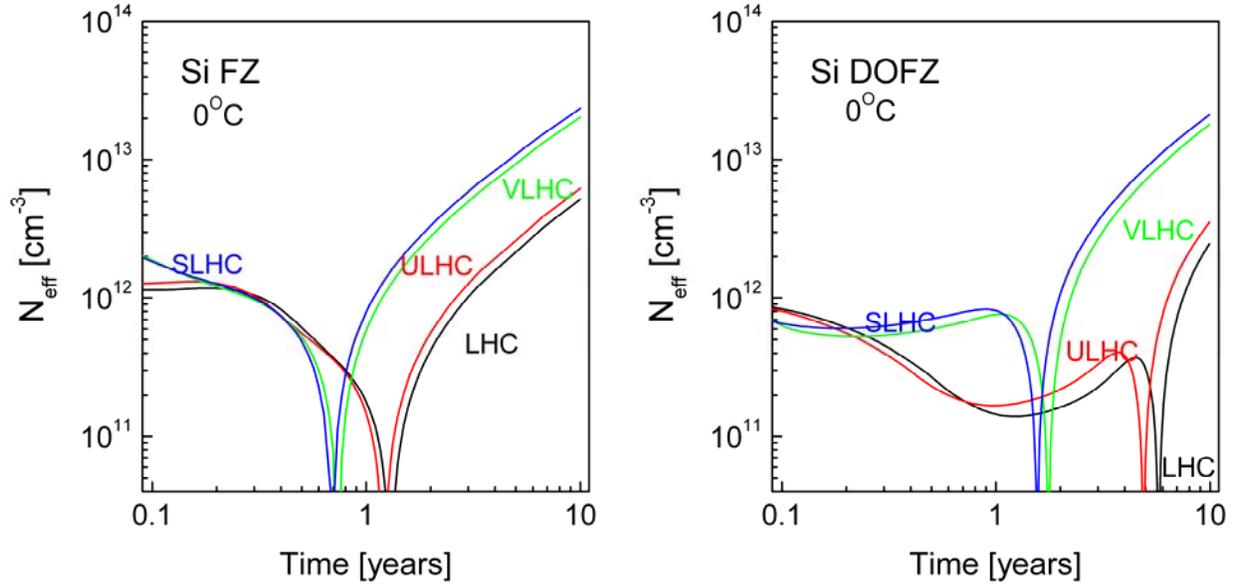

Fig. 4.
Time dependence of $N_{eff}$ for FZ and DOFZ detectors at $0^O C$ at future HEP facilities.

In the hypothesis considered, the model is able to explain and predict the main characteristics of the degradation of detectors starting from microscopic phenomena.

**Acknowledgments**

This work has partially been supported by the Romanian Scientific Programmes CERES and MATNANTECH under contracts C4-69 and 219 (404).